\documentclass[twocolumn,secnumarabic,amssymb, amsmath, nofootinbib,tightenlines,
nobibnotes, aps, prl,epsfig]{revtex4}
\usepackage{graphicx}
\usepackage{dcolumn}
\usepackage{bm}
\begin{document}
\preprint{APS/123-QED}
\title{An Evaluation of The Proton Structure Functions $F_{2}$ and $F_{L}$ at Small $x$ }

\author{G.R.Boroun}%
 \email{grboroun@gmail.com; boroun@razi.ac.ir }
 \author{B.Rezaei }
\altaffiliation{brezaei@razi.ac.ir}
\affiliation{ Physics Department, Razi University, Kermanshah
67149, Iran}

\date{\today}
\begin{abstract}
We describe the determination of the DIS structure functions
$F_{2}$ and $F_{L}$ by using the singlet
Dokshitzer-Gribov-Lipatov-Altarelli-Parisi (DGLAP)  and
Altarelli-Martinelli equations at small values of $x$. The
determination of the longitudinal structure function is presented
as a parameterization of $F_{2}(x,Q^{2})$ and its derivative.
Analytical expressions for $\sigma_{r}(x,Q^{2})$ in terms of the
effective parameters of the parameterization of $F_{2}(x,Q^{2})$
and $F_{L}(x,Q^{2})$ are presented. This analysis is enriched by
including the higher-twist effects in calculation of the reduced
cross sections which is important at low-$x$ and low-$Q^{2}$
regions. Numerical calculations and comparison with H1 data
demonstrate that the suggested method provides reliable
$F_{L}(x,Q^{2})$ and $\sigma_{r}(x,Q^{2})$ at low $x$ in a wide
range of the low absolute four-momentum transfers squared
($1.5~\mathrm{GeV}^{2}<Q^{2}<120~\mathrm{GeV}^{2}$) at moderate
and high inelasticity. Expanding the method to low and ultra low
values of $x$ can be considered in the process analysis of new
colliders. We compare the obtained longitudinal structure function
with respect to the LHeC simulated uncertainties
[CERN-ACC-Note-2020-0002, arXiv:2007.14491 [hep-ex] (2020)] with
the results from CT18 [Phys.Rev.D{\bf103}, 014013(2021)]
parametrization
model.\\

\end{abstract}
 \pacs{***}
\maketitle
\subsection{I. INTRODUCTION}
The experimental determination of the longitudinal structure
function is a realistic prospect at high energy electron-proton
colliders. First measurements of $F_{L}$
  at small $x$ were performed at HERA [1]. A next
  generation of ep colliders is under design, the Large Hadron
electron Collider (LHeC) [2,3] and
  the Future Circular Collider electron- hadron (FCC-eh) [4]
  where these measurements can be performed with
  much increased precision and extended to much lower values of $x$
  and high $Q^2$. The electron-proton center-of-mass energy at the LHeC can reach to $\sqrt{s} \simeq 1.3~
\mathrm{TeV}$, which this is about 4 times the center-of-mass
energy range of ep collisions at HERA [2,3]. The LHeC is designed
to become the finest new microscope for exploring new physics, as
the kinematic range in the ($x,Q^{2}$) plane for electron and
positron neutral-current (NC) in the perturbative region is well
below $x{\approx}10^{-6}$ and extends up to $Q{\simeq}1~
\mathrm{TeV}$. HERA has also reached to $x{\sim}10^{-6}$ in
$Q^{2}$ values $0.2$, $0.11$ and $0.045~\mathrm{GeV}^{2}$ which is
related  to the H1 svx-mb, ZEUS BPC and ZEUS BPT data sets
respectively [4]. This behavior will be extended down to $x \simeq
10^{-7}$ at the FCC-eh option of a Future Circular Collider
program [5]. The FCC-eh collider would reach a center-of-mass
energy of $\sqrt{s}=3.5~\mathrm{TeV}$ at a similar luminosity as
the LHeC. Deep inelastic scattering measurements at the FCC-eh and
the LHeC will allow the determination of parton distribution
functions at very small $x$ as they are pertinent in
investigations of lepton-hadron processes in ultra-high energy
(UHE) neutrino astroparticle physics [5]. Moreover a similar very
high energy electron-proton/ion collider (VHEep) [6] has been
suggested based on plasma wakefield acceleration, albeit with very
low luminosity. The center-of-mass energy, in this collider, is
close to $10~\mathrm{TeV}$ which is relevant in investigations of
new strong interaction dynamics related to high-energy cosmic rays
and gravitational physics (the luminosity estimate is about
six orders of magnitude below that of LHeC).\\
Recently several methods for the determination of the longitudinal
structure function in the nucleon from the proton structure
function have been proposed [7-11]. The method is based on a form
of the deep inelastic lepton-hadron scattering (DIS) structure
function which was proposed by  Block-Durand-Ha (BDH) in Ref.[12].
This new parameterization  describes accurately the results for
the high energy ep and isoscalar ${\nu}N$ total cross sections.
These cross sections obey an analytic expression as a function of
${\mathrm{\ln}}\mathrm{E}$ at large energies $\mathrm{E}$ of the
incident particle. Indeed, this parameterization is  relevant for
investigations of ultra-high energy processes, such as scattering
of cosmic
neutrinos from hadrons [7].\\
At low values of $x$, the transversal structure function
$F_{2}(x,Q^{2})$ and the longitudinal structure function
$F_{L}(x,Q^{2})$ are defined solely via the singlet quark
$xf_{s}(x,Q^{2})$ and gluon density $xf_{g}(x,Q^{2})$ as
\begin{eqnarray}
F_{k}(x,Q^{2})=<e^{2}>\sum_{a=s,g}\bigg{[}B_{k,a}(x){\otimes}xf_{a}(x,Q^{2})
\bigg{]},~~~k=2,L \nonumber
\end{eqnarray}
where $<e^{2}>=\frac{\sum_{i=1}^{N_{f}} e_{i}^{2}}{N_{f}}$ is the
average  charge squared for ${N_{f}}$ which ${N_{f}}$ denotes the
number of effective massless flavours. The quantities $B_{k,a}(x)$
are the known Wilson coefficient functions and the parton
densities fulfil the renormalization group evolution equations.
Here the non-singlet densities become negligibly small in
comparison with the singlet densities. The symbol $\otimes$
indicates convolution over the variable $x$ by the usual form,
$f(x){\otimes}g(x)=\int_{x}^{1}
\frac{dz}{z}f(z,\alpha_{s})g(x/z)$. It is important to resum the
leading $\alpha_{s}\log(1/x)$ contributions at low values of $x$,
where this resummation is accomplished by the BFKL equation [13].
In this region, the gluon density is predicted to increase as this
singular behavior of growth in $x$ is the characteristic property
of the BFKL gluon density.\\
Some time ago a proposal was published to look for the
longitudinal and transversal structure functions in deep inelastic
scattering (DIS) [14].  The authors in Ref.[14] showed that it is
possible to obtain scheme independent evolution equations for the
structure functions by the following form
\begin{eqnarray}
\frac{{\partial}F_{2}(x,Q^{2})}{{\partial}{\ln}Q^{2}}\sim
\Gamma_{22}\otimes F_{2}(x,Q^{2})+\Gamma_{2L}\otimes
F_{L}(x,Q^{2}),
\end{eqnarray}
\begin{eqnarray}
\frac{{\partial}F_{L}(x,Q^{2})}{{\partial}{\ln}Q^{2}}\sim
\Gamma_{L2}\otimes F_{2}(x,Q^{2})+\Gamma_{LL}\otimes
F_{L}(x,Q^{2}).
\end{eqnarray}
The method is based on physical observables, $F_{L}$ and $F_{2}$.
The anomalous dimensions $\Gamma_{ij}$ are computable in
perturbative QCD. The structure functions $F_{2}$ and $F_{L}$ are
related to the cross sections $\sigma_{T}$ and $\sigma_{L}$ for
interaction of transversely and longitudinally polarised photons
with protons. The reduced cross section for deep-inelastic
lepton-proton scattering depends on these independent structure
functions in the combination
\begin{eqnarray}
{\sigma}(x,Q^{2})=F_{2}(x,Q^{2})-\frac{y^{2}}{Y_{+}}F_{L}(x,Q^{2}),
\end{eqnarray}
where $Y_{+}=1+(1-y)^2$, $y={Q^{2}}/{xs}$ denotes the inelasticity
and $s$ stands for the center-of-mass  energy squared  of
incoming electrons and protons. As usual $x$ is the Bjorken
scaling parameter and $Q^{2}$ is the four momentum transfer in a
deep
inelastic scattering process.\\
In QCD, structure functions are defined as convolution of
universal parton momentum distributions inside the proton and
coefficient functions, which contain information about the
boson-parton interaction [15,16]. The standard and the basic tools
for theoretical investigation of DIS structure functions are the
DGLAP evolution equations [17,18]. The DGLAP equations based on
the parton model and perturbative QCD theory successfully and
quantitatively interpret the $Q^{2}$-dependence of parton
distribution functions (PDFs). It is so successful that most of
the PDFs are extracted by using the DGLAP equations up to now.
These equations can be used to extract the deep inelastic
scattering structure functions
of proton.\\
 The longitudinal structure
function $F_{L}(x,Q^{2})$ of the proton in terms of coefficient
function is given by [17]
\begin{eqnarray}
x^{-1}F_{L}(x,Q^{2})&=&C_{L,ns}(\alpha_{s},x){\otimes}q_{ns}(x,Q^{2})\nonumber\\
&&+<e^{2}>[C_{L,s}(\alpha_{s},x){\otimes}q_{s}(x,Q^{2})\nonumber\\
&&+C_{L,g}(\alpha_{s},x){\otimes}g(x,Q^{2})],
\end{eqnarray}
where $q_{ns}$, $q_{s}$ and $g$ are the flavour non singlet,
flavour singlet and gluon density respectively. The coefficient
functions $C_{L,a}(a=q,g)$ can be written in a perturbative
expansion as follows [19]:
\begin{eqnarray}
C_{L,a}(\alpha_{s},x)=\sum_{n=1}(\frac{\alpha_{s}}{4\pi})^{n}c^{(n)}_{L,a}(x),\nonumber
\end{eqnarray}
where $n$ denotes the order in running coupling. The coupled DGLAP
evolution equations for the singlet quark structure function
$F_{s}(x,Q^{2})=\sum_{i}x[q_{i}(x,Q^{2})+\overline{q}_{i}(x,Q^{2})]$
and the gluon distribution $G(x,Q^{2})=xg(x,Q^{2})$ can be written
as
\begin{eqnarray}
\frac{{\partial}G(x,Q^{2})}{{\partial}{\ln}Q^{2}}&=&P_{gg}(\alpha_{s},x){\otimes}
G(x,Q^{2})\nonumber\\
&&+P_{gq}(\alpha_{s},x){\otimes}F_{s}(x,Q^{2}),
\end{eqnarray}
\begin{eqnarray}
\frac{{\partial}F_{s}(x,Q^{2})}{{\partial}{\ln}Q^{2}}&=&P_{qq}(\alpha_{s},x){\otimes}
F_{s}(x,Q^{2})\nonumber\\
&&+2N_{f}P_{qg}(\alpha_{s},x){\otimes}G(x,Q^{2}).
\end{eqnarray}
The splitting functions $P_{ij}$ are the Altarelli-Parisi kernels
at LO up to  high-order corrections  [20]
\begin{eqnarray}
P_{ij}(\alpha_{s},x)=\sum_{n=1}(\frac{\alpha_{s}}{4\pi})^{n}P^{(n)}_{ij}(x),\nonumber
\end{eqnarray}
where $\alpha_{s}(Q^{2})$ is the running coupling.\\
The main purpose of the article is to study the relationship
between the structure functions, which is expected to be more
reliable at small $x$. Indeed, with respect to the DGLAP evolution
equations, the direct relationship between the longitudinal
structure function and the proton structure function is examined.
Then the
 effects of Higher-twist (HT) on the reduced cross sections at
 low-$Q^{2}$ values are considered. The organization of this
paper is as follows. In section II we introduce the basic formula
used for the definition of decoupling DGLAP evolution equations
into the proton and longitudinal structure functions. In section
III we present the longitudinal structure function with respect to
the parameterization of $F_{2}$. In section IV we also present the
effective exponent for the singlet structure function in an
independent method. Finally the formalism of HT effects used in
this analysis in section V is described.  The main results and
finding of the present longitudinal structure function and reduced
cross section at moderate and high inelasticity are discussed in
detail in section VI. In the same section, we present the
extracted HT effects at low $Q^{2}$ values and show detailed
comparisons with the experimental data. We also expand the
available energy to the range of new collider energies (i.e., LHeC
and FCC-eh). This section also includes a brief discussion of the
implication of the finding for future research. Conclusions and
summary are summarized on Sec.VII.\\

\subsection{II. BASIC FORMULA}

The standard parameterization of the singlet and gluon
distribution functions for $x{\rightarrow}0$, is given by [21,22]
\begin{eqnarray}
F_{2}^{s}(x,Q^{2})_{x{\rightarrow}0}&=&A_{s}(Q^{2})x^{-\lambda_{s}(Q^{2})},\nonumber\\
G(x,Q^{2})_{x{\rightarrow}0}&=&A_{g}(Q^{2})x^{-\lambda_{g}(Q^{2})},
\end{eqnarray}
where $A_{s}$ and $A_{g}$ are $Q^{2}$ dependent and $\lambda^{,}$s
are strictly positive. This behavior of the singlet structure
function was proposed by Lopez and Yndurain [21], and the
inclusive electroproduction on a proton was studied at low $x$ and
low $Q^{2}$ using a soft and hard Pomeron in Ref.[22]. The gluon
exponent at low values of $x$ at $Q^{2}=1~\mathrm{GeV}^{2}$
obtained by MSTW08 NLO was $0.428^{+0.066}_{-0.057}$ [23]. The
effective exponent for the gluon distribution at $Q^{2}=10
~\mathrm{GeV}^{2}$ and $x=10^{-4}$  obtained by NNPDF3.0, CT14,
MMHT14, ABM12 and CJ15 had values of  0.20, 0.15, 0.29, 0.15 and
0.14 respectively. The value obtained by fixed coupling LLx BFKL
gives $\lambda_{g}{\simeq}0.5$, which is the so-called
hard-Pomeron exponent. Some other phenomenological models have
also been proposed for the singlet structure function exponents in
Refs.[24] and [25]. The singlet and gluon exponents are determined
and applied to the deep inelastic lepton nucleon scattering at low
values of $x$ in Ref.[26]. Ref.[27] used a form inspired by double
asymptotic $x,Q^{2}$ scaling. The hard Pomeron behavior of the
photon-proton cross section based on a simple power-law behavior
and double asymptotic scaling at low $x$ values for
$10<W<10^{4}~\mathrm{GeV}$ in the kinematic range of VHEep is
shown in Ref.[6]. Recently, authors in Ref.[28] presented a
tensor-Pomeron model where it is applied to low-$x$ deep inelastic
lepton-nucleon scattering and photoproduction processes. In this
model, in addition to the soft tensor Pomeron, a hard tensor
Pomeron and Reggeon exchange included. In this case the
hard-Pomeron intercept was determined  to $0.3008(^{+73}_{-84})$
with the latest HERA data for $x<0.01$. An effective behavior for
the singlet structure function is reported in Refs.[24] and [29].
This effective exponent was found to be independent of $x$ and to
increase linearly with ${\ln}Q^{2}$. Indeed the function
$\lambda(Q^{2})$ was determined from fits of the form
$F_{2}(x,Q^{2})=c(Q^{2})x^{-\lambda(Q^{2})}$ to the H1 data, and
the coefficients $c(Q^{2})$ are approximately independent of
$Q^{2}$ and $\lambda(Q^{2})$ rises linearly with ${\ln}Q^{2}$.\\
Now we use the hard pomeron behavior for the distribution
functions (i.e., Eq.(7)) in Eqs.(4-6) and obtain the modified
evolution equations, which take into account all the modifications
mentioned above, by the following forms
\begin{eqnarray}
\frac{{\partial}G(x,Q^{2})}{{\partial}{\ln}Q^{2}}&=&G(x,Q^{2})\Phi_{gg}(x,Q^{2})\nonumber\\
&&+F_{2}(x,Q^{2})\Theta_{gq}(x,Q^{2}),
\end{eqnarray}
\begin{eqnarray}
\frac{{\partial}F_{2}(x,Q^{2})}{{\partial}{\ln}Q^{2}}&=&F_{2}(x,Q^{2})\Phi_{qq}(x,Q^{2})\nonumber\\
&&+G(x,Q^{2})\Theta_{qg}(x,Q^{2}),
\end{eqnarray}
and
\begin{eqnarray}
F_{L}(x,Q^{2})&=&F_{2}(x,Q^{2})I_{L,q}(x,Q^{2})\nonumber\\
&&+G(x,Q^{2})I_{L,g}(x,Q^{2}).
\end{eqnarray}
The convolution and the compact form of the kernels
 are given by
\begin{eqnarray}
\Phi_{qq}(x,Q^{2})&=&P_{qq}(x,\alpha_{s}){\odot} x^{\lambda_{s}(Q^{2})}\nonumber\\
\Phi_{gg}(x,Q^{2})&=&~P_{gg}(x,\alpha_{s}){\odot}
x^{\lambda_{g}(Q^{2})}\nonumber\\
\Theta_{gq}(x,Q^{2})&=&\frac{18}{5}P_{gq}(x,\alpha_{s}){\odot}
x^{\lambda_{s}(Q^{2})}\nonumber\\
\Theta_{qg}(x,Q^{2})&=&\frac{10N_{f}}{18}P_{qg}(x,\alpha_{s}){\odot}
x^{\lambda_{g}(Q^{2})}\nonumber\\
I_{L,q}(x,Q^{2})&=&~\frac{18}{5N_{f}}C_{L,q}(x,\alpha_{s}){\odot}
x^{\lambda_{s}(Q^{2})}\nonumber\\
I_{L,g}(x,Q^{2})&=&<e^{2}>C_{L,g}(x,\alpha_{s}){\odot}
x^{\lambda_{g}(Q^{2})},
\end{eqnarray}
where we have defined the convolution form to be
\begin{eqnarray}
f(x){\odot}g(x)=\int_{x}^{1}\frac{dz}{z}f(\alpha_{s}, z)g(z).
\end{eqnarray}
These equations (i.e., Eqs.(8-10)) can now be easily decoupled.
The idea is to modify the evolution equations in order to satisfy
simultaneously the decoupled evolution equation based on the
structure functions. Indeed, in this method, we separate the gluon
distribution function from Eqs.(8-10). In fact the singlet and
longitudinal structure functions contain the gluon distribution
which comes from the perturbative QCD. Therefore the solution of
Eq.(10) is straightforward and given by
\begin{eqnarray}
G(x,Q^{2})=\frac{F_{L}(x,Q^{2})}{I_{L,g}(x,Q^{2})}-F_{2}(x,Q^{2})\frac{I_{L,q}(x,Q^{2})}{I_{L,g}(x,Q^{2})}.
\end{eqnarray}
Using Eq.(13) in (9), our solution takes the form
\begin{eqnarray}
\frac{{\partial}F_{2}(x,Q^{2})}{{\partial}{\ln}Q^{2}}&=&\Gamma_{22}(x,Q^{2})F_{2}(x,Q^{2})\nonumber\\
&&+\Gamma_{2L}(x,Q^{2})F_{L}(x,Q^{2}),
\end{eqnarray}
where
\begin{eqnarray}
\Gamma_{22}(x,Q^{2})&=&\Phi_{qq}(x,Q^{2})-\Theta_{qg}(x,Q^{2})\frac{I_{L,q}(x,Q^{2})}{I_{L,g}(x,Q^{2})},\nonumber\\
\Gamma_{2L}(x,Q^{2})&=&\frac{\Theta_{qg}(x,Q^{2})}{I_{L,g}(x,Q^{2})}.
\end{eqnarray}
Also substituting Eq.(13) in Eq.(8) we get
\begin{eqnarray}
\frac{{\partial}F_{L}(x,Q^{2})}{{\partial}{\ln}Q^{2}}&=&\Gamma_{LL}(x,Q^{2})F_{L}(x,Q^{2})\nonumber\\
&&+\Gamma_{L2}(x,Q^{2})F_{2}(x,Q^{2}),
\end{eqnarray}
where
\begin{eqnarray}
\Gamma_{LL}(x,Q^{2})&=&T_{LL}(x,Q^{2})+\frac{{\partial}}{{\partial}{\ln}Q^{2}}{\ln}I_{L,g}(x,Q^{2}),\nonumber\\
\Gamma_{L2}(x,Q^{2})&=&T_{L2}(x,Q^{2})+I_{L,q}(x,Q^{2})\nonumber\\
&&{\times}\frac{{\partial}}{{\partial}{\ln}Q^{2}}{\ln}\frac{I_{L,q}(x,Q^{2})}{I_{L,g}(x,Q^{2})}.
\end{eqnarray}
and
\begin{eqnarray}
T_{LL}(x,Q^{2})&=&\Phi_{gg}(x,Q^{2})+\Theta_{qg}(x,Q^{2})\frac{I_{L,q}(x,Q^{2})}{I_{L,g}(x,Q^{2})},\nonumber\\
T_{L2}(x,Q^{2})&=&I_{L,q}(x,Q^{2})[\Phi_{qq}(x,Q^{2})-\Phi_{gg}(x,Q^{2})\nonumber\\
&&-\Theta_{qg}(x,Q^{2})\frac{I_{L,q}(x,Q^{2})}{I_{L,g}(x,Q^{2})}]\nonumber\\
&&+\Theta_{gq}(x,Q^{2})I_{L,g}(x,Q^{2}).
\end{eqnarray}
Therefore the evolution equations for the structure functions
$F_{2}(x,Q^{2})$ and $F_{L}(x,Q^{2})$ read as
\begin{eqnarray}
\frac{{\partial}}{{\partial}{\ln}Q^{2}} \left(\begin{array}{c}
 F_{2}(x,Q^{2}) \\
 F_{L}(x,Q^{2})
 \end{array}
 \right)=\left(
 \begin{array}{cc}
 \Gamma_{22} & \Gamma_{2L}\\
 \Gamma_{L2} & \Gamma_{LL}
 \end{array}
 \right){\times}
\left(\begin{array}{c}
 F_{2}(x,Q^{2}) \\
 F_{L}(x,Q^{2})
 \end{array}
 \right).\nonumber\\
\end{eqnarray}

\subsection{III. DETERMINING THE LONGITUDINAL STRUCTURE FUNCTION}

One can rewrite Eq.(13)  related to the proton structure function
$F_{2}(x,Q^{2})$ and its derivative with respect to ${\ln}Q^{2}$,
as we have
\begin{eqnarray}
F_{L}(x,Q^{2})&=&\frac{1}{\Gamma_{2L}}\frac{{\partial}F_{2}(x,Q^{2})}{{\partial}{\ln}Q^{2}}-\frac{\Gamma_{22}}{\Gamma_{2L}}F_{2}(x,Q^{2}).
\end{eqnarray}
This relation (i.e., Eq.(20)) help to estimate the proton
longitudinal structure function in terms of the effective
parameters of the parametrization of $F_{2}(x,Q^{2})$. We now
employ the $F_{2}$ parameterization of Ref.[12] which is obtained
from a combined fit of the HERA data [4] with $x\leq 0.1$ and $W
\geq 25~ \mathrm{GeV}$ and use it in Eq.(20). This parametrization
describes fairly good the available experimental data on the
proton structure function in agreement with the Froissart [30]
bound behavior. The explicit expression for the $F_{2}$
parametrization [12] is given by the following form
\begin{eqnarray}
F^{\gamma p}_{ 2}(x,Q^{2})& =& D(Q^{2})(1-
x)^{n}\sum_{m=0}^{2}A_{m}(Q^{2})L^{m},
\end{eqnarray}
where
\begin{eqnarray}
A_{0}(Q^{2})& =& a_{00} + a_{01}
{\ln}(1+\frac{Q^{2}}{\mu^{2}}),\nonumber\\
 A_{1}(Q^{2})& =& a_{10} + a_{11} {\ln}(1+\frac{Q^{2}}{\mu^{2}}) + a_{12}{\ln}^{2}(1+\frac{Q^{2}}{\mu^{2}})
 ,\nonumber\\
A_{2}(Q^{2})& =& a_{20} + a_{21} {\ln}(1+\frac{Q^{2}}{\mu^{2}}) +
a_{22}{\ln}^{2}(1+\frac{Q^{2}}{\mu^{2}})
 ,\nonumber\\
D(Q^{2})& =& \frac{Q^{2}(Q^{2}+\lambda M^{2})}{(Q^{2}+M^{2})^2},\nonumber\\
L^{m}&=&\ln^{m}(\frac{1}{x}\frac{Q^{2}}{Q^{2}+\mu^{2}}).\nonumber
\end{eqnarray}
Here $M$ and $\mu^{2}$ are the effective mass  a scale factor
respectively. The fixed  parameters are defined by the
Block-Halzen [31] fit to the real photon-proton cross section. The
additional parameters with their statistical errors are given in
Table I. Eventually, inserting Eq.(21) and its derivative in (20)
one obtains
\begin{eqnarray}
F_{L}(x,Q^{2})&=&\mathcal{R}(\digamma_{i,j}(x,Q^{2}))D(Q^{2})(1-
x)^{n}\nonumber\\
&&{\times}\sum_{m=0}^{2}A_{m}(Q^{2})L^{m},
\end{eqnarray}
where
\begin{eqnarray}
\mathcal{R}(\digamma_{i,j}(x,Q^{2}))&=&\frac{1}{\Gamma_{2L}}\{\frac{{\partial}{\ln}D(Q^{2})}{{\partial}{\ln}Q^{2}}\\
&&+\frac{{\partial}{\ln}(\sum_{m=0}^{2}A_{m}(Q^{2})L^{m})}{{\partial}{\ln}Q^{2}}\}
-\frac{\Gamma_{22}}{\Gamma_{2L}}.\nonumber
\end{eqnarray}
We reiterate that this analysis for the longitudinal structure
function behavior is based  on the proton structure function and
its derivative. The longitudinal structure function
$F_{L}(x,Q^{2})$ is obtained according to the parameterization
known for the proton structure function $F_{2}(x,Q^{2})$ where
determined from the existing experimental data. The obtained
expression for the parameterization of $F_{L}(x,Q^{2})$ has the
same behavior as Eq.(21) with the Froissart boundary condition
which evaluated in terms of the $\mathcal{R}$
function.\\

\subsection{IV. DETERMINING THE RATIO ${\sigma_{r}}/{F_{2}}$ }

Considering the expressions (3) and (22), we can estimate the
ratio ${\sigma_{r}}/{F_{2}}$ due to the longitudinal structure
function behavior. This ratio is given by
\begin{eqnarray}
\frac{\sigma_{r}(x,Q^{2})}{F_{2}(x,Q^{2})}&=&1-\Delta(x,y,Q^{2}),
\end{eqnarray}
where
\begin{eqnarray}
\Delta(x,y,Q^{2})&=&f(y)\mathcal{R}(\digamma_{i,j}(x,Q^{2})).
\end{eqnarray}
Here $f(y)=\frac{y^{2}}{Y_{+}}$ and $\mathcal{R}$ is defined as
\begin{eqnarray}
\mathcal{R}(\digamma_{i,j}(x,Q^{2}))&=&\frac{1}{\Gamma_{2L}}\frac{{\partial}{\ln}F_{2}(x,Q^{2})}{{\partial}{\ln}Q^{2}}
-\frac{\Gamma_{22}}{\Gamma_{2L}}.
\end{eqnarray}
The ratio ${\sigma_{r}}/{F_{2}}$ can be obtained from the
parametrization and hard-pomeron behavior of $F_{2}(x,Q^{2})$. In
Eq.(23) we considered the $\mathcal{R}$ function dependent on the
parameterization of $F_{2}(x,Q^{2})$. Now it is interesting to
confront the $\mathcal{R}$ function obtained from the effective
exponent with the properties of the power-like behavior of the
proton structure function. Let us use this behavior for evolution
of $\frac{{\partial}{\ln}F_{2}(x,Q^{2})}{{\partial}{\ln}Q^{2}}$ in
accordance with the effective exponents [24,29,33], as we have
\begin{eqnarray}
\frac{{\partial}{\ln}F_{2}(x,Q^{2})}{{\partial}{\ln}Q^{2}}\simeq
-\frac{{d}\lambda(Q^{2})}{{d}{\ln}Q^{2}}{\ln}x,
\end{eqnarray}
where the other coefficients (i.e., $A_{s}(Q^{2})$ or $c(Q^{2})$)
are approximately independent of $Q^{2}$. Therefore the
$\mathcal{R}$ function with respect to the effective exponent
behavior is found to be
\begin{eqnarray}
\mathcal{R}(\digamma_{i,j}(x,Q^{2}))&=&\frac{1}{\Gamma_{2L}}[-\frac{{d}\lambda(Q^{2})}{{d}{\ln}Q^{2}}{\ln}x]-\frac{\Gamma_{22}}{\Gamma_{2L}}.
\end{eqnarray}
In order to find the ratio ${\sigma_{r}}/{F_{2}}$, we use the
$F_{2}$ parameterization and effective exponent methods and obtain
the following forms respectively
\begin{eqnarray}
\frac{\sigma_{r}(x,Q^{2})}{F_{2}(x,Q^{2})}&=&1-\frac{y^{2}}{Y_{+}}{\bigg\{}\frac{1}{\Gamma_{2L}}[\frac{{\partial}{\ln}D(Q^{2})}{{\partial}{\ln}Q^{2}}\\
&&+\frac{{\partial}{\ln}(\sum_{m=0}^{2}A_{m}(Q^{2})L^{m})}{{\partial}{\ln}Q^{2}}]
-\frac{\Gamma_{22}}{\Gamma_{2L}}{\bigg\}}\nonumber
\end{eqnarray}
and
\begin{eqnarray}
\frac{\sigma_{r}(x,Q^{2})}{F_{2}(x,Q^{2})}=1-\frac{y^{2}}{Y_{+}}{\bigg\{}\frac{1}{\Gamma_{2L}}[-\frac{{d}\lambda(Q^{2})}{{d}{\ln}Q^{2}}{\ln}x]-\frac{\Gamma_{22}}{\Gamma_{2L}}{\bigg\}}.
\end{eqnarray}
Due to the positivity of the cross sections for longitudinally and
transversely polarized photons scattering off protons, the
$\mathcal{R}$ function obey the relation
$0{\leq}\mathcal{R}{\leq}1$. Thus the contribution of the
$\mathcal{R}$ function to the cross section can be sizable due to
the coefficient functions with respect to the parameterization and
effective exponent methods. An important advantage of this method
can be used to determine the difference between the measured
$\sigma_{r}$ and the extrapolated $F_{2}$ in colliders.\\

\subsection{V. HIGHER TWIST CORRECTION}

In a wide kinematic region in terms of $x$ and $Q^{2}$, one can
describe the deeply inelastic structure functions using
leading-twist  corrections in QCD. At low $Q^{2}$ values there are
constraints on the structure function $F_{i}(x,Q^{2})$ which
follow from eliminating the kinematical singularities at $Q^{2}=0$
from the hadronic tensor $W^{\mu{\nu}}$. In this region the higher
twists (HT) concept is introduced, in which the operator product
expansion leads to the representation [14]
\begin{eqnarray}
F_{2}(x,Q^{2})=\sum_{n=0}^{\infty}\frac{C_{n}(x,Q^{2})}{(Q^{2})^{n}},
\end{eqnarray}
where the function $C_{0}$ refers to as leading twist (LT) and
$C_{\geq{1}}$ refers to as higher twist (HT). Higher twist
corrections  emerge both in the region of large and small values
of $x$. These corrections arise from the struck proton$^{,}$s
interaction with target remnants reflecting confinement. The
introduction of higher-twist terms is one possible way to extend
the DGLAP framework to low $Q^{2}$ values. Indeed the conventional
DGLAP evolution dose not describe the DIS data in the low
$x-Q^{2}$ region very well [33]. The higher-twist effects are
parameterized in the form of a phenomenological unknown function,
and the values of the unknown parameters are obtained from fits to
the experimental data [34,35]. It is customary to correct the
leading-twist structure function by adding a sentence that is
inversely related to $Q^{2}$ as the phenomenological power
correction to the structure function from the HT effects
 is considered by the following form
 \begin{eqnarray}
F_{2}(x,Q^{2})=F_{2}^{LT}(x,Q^{2})(1+\frac{C_{HT}(x)}{Q^{2}}),
 \end{eqnarray}
where the $F_{2}^{LT}$ is the leading twist contribution to the
structure function $F_{2}$ and the higher-twist coefficient
function $C_{HT}(x)$ is determined from fit to the data. Here
$C_{HT}=0.12~\pm~0.07~\mathrm{GeV}^{2}$ is the result of a fit
[34,35]. Note that in such a parametrization the power correction
does not depend on $x$. Some  authors have reported [36] the HT
coefficient function parameterized  as follows
\begin{eqnarray}
C_{HT}(x)=h_{0}(h_{2}(x)x^{h_{1}}+\gamma),\nonumber
\end{eqnarray}
which the parameters represent the HT effects in the perturbative
QCD. The corresponding parameters obtained from the QCD analysis
fit with HT effects included at the initial scale $Q_{0}^{2}$.\\
In the following we study on the determination of the higher twist
contributions in deeply-inelastic structure function [37,38]. To
better illustrate our analysis for the longitudinal structure
function at low $Q^{2}$ values, we added a higher twist term in
the description of the parameterization of $F_{2}(x,Q^{2})$ [12].
To elaborate further, $F_{2}$ can be expressed in terms of the
higher twist coefficients as
\begin{eqnarray}
F_{2}=D(Q^{2})(1-
x)^{n}(1+\frac{C_{HT}}{Q^{2}})\sum_{m=0}^{2}A_{m}(Q^{2})L^{m}.
\end{eqnarray}
Now we present an analytical analysis equation for the
longitudinal structure function with considering the HT effects.
The phenomenological form for the HT effects in the obtained
longitudinal structure function is considered as follows
\begin{eqnarray}
F_{L}(x,Q^{2})&=&\mathcal{R}_{HT}(\digamma_{i,j}(x,Q^{2}))D(Q^{2})(1-
x)^{n}\nonumber\\
&&\sum_{m=0}^{2}A_{m}(Q^{2})L^{m},
\end{eqnarray}
where
\begin{eqnarray}
\mathcal{R}_{HT}(\digamma_{i,j}(x,Q^{2}))&=&\frac{1}{\Gamma_{2L}}(1+\frac{C_{HT}}{Q^{2}})\{\frac{{\partial}{\ln}D(Q^{2})}{{\partial}{\ln}Q^{2}}\nonumber\\
&&+\frac{{\partial}{\ln}(\sum_{m=0}^{2}A_{m}(Q^{2})L^{m})}{{\partial}{\ln}Q^{2}}\}\nonumber\\
&&-\frac{\Gamma_{22}}{\Gamma_{2L}}
-\frac{C_{HT}}{Q^{2}}\frac{1+\Gamma_{22}}{\Gamma_{2L}}.
\end{eqnarray}
We conclude that in this method  we consider the higher twist
effects in the longitudinal structure function behavior. As a
result of this study, we will show that such corrections are
sizable at small region of $Q^{2}$. Hence, it will be interesting
to see  the significant change in the longitudinal structure
function and the reduced cross section after including the HT corrections.\\

\subsection{VI. RESULTS AND DISCUSSIONS}

In this section, we present our results that have been obtained
for the longitudinal structure function $F_{L}(x,Q^{2})$ and
reduced cross section $\sigma_{r}(x,Q^{2})$ from data mediated by
the parameterization of $F_{2}(x,Q^{2})$. Then we present the
results obtained for $F_{L}(x,Q^{2})$ and $\sigma_{r}(x,Q^{2})$
with and without considering the HT effects. We have calculated
the $Q^{2}$-dependence of the longitudinal structure function and
reduced cross section with respect to the standard representations
for the QCD couplings [39-41] at low values of $x$. The results
for the longitudinal structure function  are presented in Fig.1
and compared with the H1 data [1] as accompanied with total
errors. Calculations have been performed at fixed values of the
singlet and gluon exponents as they are controlled by Pomeron
exchange [42,43]. In this figure (i.e., Fig.1) and the rest, the
error bands represent the upper and lower limit of the uncertainty
shown in Table I. As can be seen in this figure (i.e., Fig.1), the
results are comparable with the H1 data in  the interval
$1.5~\mathrm{GeV^{2}}< Q^{2}< 120~\mathrm{GeV^{2}}$. At all
$Q^{2}$ values,  the extracted longitudinal structure functions
are in  good agreement with the simulated data.\\
 In Fig.2 our calculations for the longitudinal structure
function are associated with  the LHeC simulated uncertainties
[3]. These simulated uncertainties for $F_{L}$ measurement
recently published by the LHeC study group and reported by Ref.
[3]. Indeed in Fig.2 a comparison between the statistical errors
described in the parameterization method and the simulated
uncertainties for the longitudinal structure function  is shown
along with the results of the central $F_{L}$. These comparisons
are shown in the interval $2.5 ~
\mathrm{GeV}^{2}{\leq}Q^{2}{\leq}2000~\mathrm{GeV}^{2}$ as
described in [3]. In order to present more detailed discussions on
our findings, we also compare the results for the longitudinal
structure function with CT18 [44] in Fig.3. As can be seen from
the related figures, the longitudinal structure function results
are consistent with the CT18 NLO at moderate and large values of
$Q^{2}$. These results are comparable to the CT18 NLO results and
different from NNLO results at moderate $Q^{2}$ values. This is
predictable because for CCFR and NuTeV kinematics, the difference
between the NNLO results of the fixed flavor number scheme (FFNS)
and each variable flavor number scheme (VFNS) is expected to be
significant less than the accuracy of experimental data. Indeed at
NNLO the gluon becomes more negative for very small values of $x$.
The fact that the gluon is not directly physical is
well-illustrated by the third-order of the longitudinal
coefficient functions. At lowest and moderate $Q^{2}$ values we
expect the NLO and NNLO predictions to be different in comparison
with the high $Q^{2}$ in some regions.\\
 In this figure (i.e.,
Fig.3) the straight lines represent the CT18 NLO and CT18 NNLO QCD
analysis and the red circles represent our results as accompanied
with the LHeC simulated uncertainties. Indeed the CT18 PDFs are an
updated version of the CT14 [45]. Other CT18 versions, such as
CT18A, Z and X, have differences in the renormalization scale and
coupling constant. Corresponding to the value of $Q^{2}$, schemes
such as general- mass variable flavor number scheme (GM-VFNS) and
zero-mass variable flavor number scheme (ZM-VFNS) are examined by
considering the production threshold of charm and bottom-quarks.
The fixed flavor number scheme is valid for
$Q^{2}{\lesssim}m^{2}_{H}$ and ZM-VFNS is valid for
$Q^{2}{\gg}m^{2}_{H}(H=c,b)$. For realistic kinematics we used the
 GM-VFNS  which is similarly to the ZM-VFNS in the $Q^{2}/m^{2}_{H}{\rightarrow\infty}$ limit for
the CT18 results.\\
We found slightly disagreements between $x$-space results
calculated from the parameterization method and the CT18 analysis
at moderate $Q^{2}$ values for the longitudinal structure
function. In fact, one of the differences is related to the shape
of the introduced distribution functions. The parton distribution
functions in the CT18 at the initial scale $Q_{0}^{2}$ are
parametrized by Bernstein polynomials multiplied by the standard
$x^{a}$ and $(1-x)^{b}$ factors that determine the small-$x$ and
large-$x$ asymptotics [44]. These polynomials are very flexible
across the whole interval $0<x<1$. While we have used a single
power law behavior for distribution functions. Another point in
the CT18 global analysis is that the starting scale $Q^{2}_{0}$
for evolution of the PDFs is around $1.69~\mathrm{GeV^{2}}$ in
comparison with the initial scale of the parameterization method
which  it is almost $0.10~\mathrm{GeV^{2}}$. Indeed the  parton
distribution functions of CT18 determined using the LHC data and
the combined HERA I+II data sets, along with the data sets
presented in the CT14 global
QCD analysis [45].\\
In Fig.4 the ratio of the longitudinal to transverse cross
sections $R(x,Q^{2})=\sigma_{L}(x,Q^{2})/\sigma_{T}(x,Q^{2})$ is
plotted in a wide range of $Q^{2}$ values. This ratio is expressed
in terms of the longitudinal-to-transverse ratio of structure
functions as defined by
\begin{eqnarray}
R(x,Q^{2})&=&\frac{F_{L}(x,Q^{2})}{F_{2}(x,Q^{2})-F_{L}(x,Q^{2})}\nonumber\\
&&=\frac{\mathcal{R}(\digamma_{i,j}(x,Q^{2}))}{1-\mathcal{R}(\digamma_{i,j}(x,Q^{2}))}\nonumber\\
&&=\left\{
 \begin{array}{cc}
\frac{ Eq.(23)}{1-Eq.(23)} & \mathrm{parameterization~ of~ F_{2}}\\\\
 \frac{ Eq.(28)}{1-Eq.(28)} & \mathrm{effective~ exponent}
 \end{array}
 \right\}\nonumber
\end{eqnarray}
In this figure we present this ratio in comparison with the color
dipole model (CDM) results [46,47,48] and experimental data
[49,50]. The value of the ratio cross sections $R$ predicted to be
$0.5$ or $0.375$ related to the color-dipole cross sections in
Refs.[46,23,48]. H1 and ZEUS collaborations in Refs.[49,50] show
that the ratio $R$ is found to be $R=0.260{\pm}0.050$ at
$3.5{\leq} Q^{2}{\leq} 45~\mathrm{GeV}^{2}$ and
$0.105^{+0.055}_{-0.037}$ at $5{\leq} Q^{2}{\leq}
110~\mathrm{GeV}^{2}$ respectively. In Fig.4, we compared our
results with the other results mentioned in the literature
[46-50]. We observe that the behavior of $R$ is very little
dependent on $x$ and $Q^{2}$ in a wide range of $x$ and $Q^{2}$
values. This behavior is observable in comparison with the
experimental data. In figures 5 and 6, the ratio of $R$  have been
depicted at fixed value of the center-of-mass energy $s$ (i.e.,
$\sqrt{s}=1.3~\mathrm{TeV}$). As can be seen in these figures, the
results are comparable with other constant values in the interval
$0.1{<}y{<}0.5$. In Fig.6 we observe that the ratio $R$ for
$\sqrt{s}=1.3~\mathrm{TeV}$ is consistent with a constant behavior
with respect to inelasticity $y$ for fixed values of $Q^{2}$.\\
 The determination of the structure function $F_{L}$ can be used
 to determine the reduced cross section $\sigma_{r}$. The $Q^{2}$-evolution results of the reduced cross section $\sigma_{r}$ are depicted
in Fig.7. These results compared with the H1 data [1] correspond
to the parameterization of $F_{2}$  [12] and the effective
exponent [24] respectively. In both methods ($F_{2}$ parameterized
and effective exponent) we find that the results are comparable
with the H1 data at fixed value of the inelasticity $y=0.49$
and at a center-of-mass energy $\sqrt{s}=225~\mathrm{GeV}$.\\
This method persuades us that the obtained results  can be
pertinent in future analysis of the ultra-high energy neutrino
data. The result of this study is shown in Fig.8. The
center-of-mass energy $\sqrt{s}=1.3~\mathrm{TeV}$ used in
accordance with the  LHeC center-of-mass energy. Also the averaged
parameter $y$ is constrained by the equality $<y>=0.49$. The
longitudinal structure function and the reduced cross section are
predicted in this energy. These results accompanied with the
statistical errors of the parameterization of $F_{2}$ at
$2.5{\leq}Q^{2}{\leq}2000~\mathrm{GeV}^{2}$ have been shown in
this figure (i.e., Fig.8). Also the difference between the central
values of the reduced cross sections due to the parameterization
of $F_{2}$ and effective exponent methods has been shown in this
figure.\\
In Fig.9, our reduced cross sections have been shown as a function
of $Q^{2}$ values at low $x$ values with and without the HT
corrections. The effect of the HT contribution to $\sigma_{r}$ can
be seen from this figure. Indeed, the HT corrections lead to large
corrections for low $Q^{2}$ values. We compare these results for
the reduced cross section with the H1 data [1] for some selected
values of low $Q^{2}$ at a center-of-mass energy
$\sqrt{s}=319~\mathrm{GeV}$ and a fixed inelasticity value
$y=0.8$. From the data versus model comparisons, the difference
between our results with and without the HT corrections can be
clearly be observed. In Fig.10 the results for the longitudinal
structure function $F_{L}(x,Q^{2})$ have been shown with and
without HT corrections at fixed value of the invariant mass W
(i.e. $W=230~ \mathrm{GeV}$) at low values of $x$. As can be seen
in this figure, the results are comparable with the H1 data [51]
at all $Q^{2}$ values. In comparison with the H1 data we observe
that the results with the HT corrections are better than the
results without the HT corrections at low values of $Q^{2}$. One
can see that the differences between theory and data are decreased
by including HT effects in the
analysis.\\

\subsection{VII. Summary}

In conclusion, we have computed the longitudinal structure
function $F_{L}$ using the parameterization of $F_{2}$ to find an
analytical solution for the DGLAP evolution equations. The
obtained explicit expression for the longitudinal structure
function is determined by parameterized the transversal proton
structure function. Then we present a further development of the
method of extraction of the reduced cross section from the
parameterization of $F_{2}$ and $F_{L}$. The
calculations are consistent with the H1 data from HERA collider.\\
In this work, we also calculated the longitudinal structure
function and reduced cross section using the effective exponent
measured by the H1 collaboration. At low $x$, the exponent
$\lambda$ has been observed to increase  linearly  with
$\ln{Q^{2}}$. In this regard, data has been collected in the
kinematic range $3.10^{-5} \leq x \leq 0.2$ and $1.5\leq Q^{2}
\leq 150~\mathrm{GeV}^{2}$. We observed that the general solutions
are in satisfactory agreements with the available experimental
data at a center-of-mass energy $\sqrt{s}=225~\mathrm{GeV}$ and a
fixed value of inelasticity $y=0.49$. Our analysis is also
enriched with the higher twist (HT) corrections to the reduced
cross section at a center-of-mass energy
$\sqrt{s}=319~\mathrm{GeV}$ and a fixed value of inelasticity
$y=0.8$, which extend to small values of $Q^{2}$. It
has been demonstrated that the HT terms are required for the low values of $Q^{2}$.\\
This persuades us that the obtained results can be extended to
high energy regime in new colliders (like in the proposed LHeC and
FCC-eh colliders). The $F_L$ measurements, as for LHeC, are much
more precise than the initial H1 measurement and extend to lower
$x$ and higher $Q^2$. These measurements will shed light on the
parton dynamics at small $x$. When confronted with the DGLAP based
predictions, one will explore the evolution dynamics deeply
and more reliably than HERA measurements did allow.\\

\subsection{ACKNOWLEDGMENTS}
 The authors are especially grateful to Max Klein for carefully
 reading the manuscript and fruitful discussions.
The authors are thankful to the Razi University for financial
support of this project. Also G.R.Boroun  thank M.Klein and
N.Armesto for allowing access to data related to simulated errors
of the longitudinal structure function at the Large Hadron
Electron Collider (LHeC). Authors would like to thank H.Khanpour
for help with preparation of the QCD parametrization
models.\\


\begin{table}[h]
\caption{ The effective parameters at low $x$ for
$0.15~\mathrm{GeV}^{2}<Q^{2}<3000~\mathrm{GeV}^{2}$ provided by
the following values. The fixed  parameters are defined by the
Block-Halzen fit to the real photon-proton cross section as
$M^{2}=0.753 \pm 0.068~ \mathrm{GeV}^{2}$ and $\mu^2 = 2.82 \pm
0.290~ \mathrm{GeV}^{2}$.}
\begin{tabular} {cccc}
\toprule \\  \multicolumn{2}{c}{parameters \quad \quad \quad ~~~~~~~~~~~~~~~~value}    \\ &&&\\ \hline \\ &&&\\
$a_{00}$& \quad  $2.550\times 10^{-1}~\pm 1.60\times10^{-2}$ \\

$a_{01}$& \quad  $1.475\times 10^{-1}~\pm 3.025\times10^{-2}$\\&&&\\

  $a_{10} $  &   \quad  $8.205\times 10^{-4}~~  \pm  4.62\times10^{-4} $  \\

  $a_{11} $  &   \quad   $-5.148\times 10^{-2}\pm 8.19\times10^{-3}$  \\

  $a_{12}$   &    \quad  $-4.725\times 10^{-3}\pm 1.01\times10^{-3}$   \\  &&&\\

 $a_{20}$   &   \quad   $2.217\times 10^{-3}\pm 1.42\times10^{-4} $ \\

 $a_{21}$   &   \quad   $1.244\times 10^{-2}\pm 8.56\times10^{-4}$  \\

 $a_{22}$    &    \quad  $5.958\times 10^{-4}\pm 2.32\times10^{-4} $ \\ &&& \\

$n$& \quad  $11.49\pm 0.99$ & &\\

$\lambda$& \quad  $2.430~\pm 0.153$ & &\\

$\chi^{2}(\mathrm{goodness~ of~ fit})$ &  \quad  $0.95$ & &\\

\hline

\end{tabular}
\end{table}

\newpage{
\section{References}
1. F.D. Aaron et al. [H1 Collaboration],
Eur.Phys.J.C\textbf{71},1579(2011); S. Chekanov et al. [ZEUS Collaboration], Phys.Lett.B{\bf 682}, 8(2009).\\
2. M.Klein,  arXiv [hep-ph]:1802.04317; M.Klein,
Ann.Phys.{\bf528}, 138(2016); N.Armesto et al.,
Phys.Rev.D{\bf100}, 074022(2019);
F.Hautmann, LHeC 2019 workshop (https://indico-cern.ch/event/835947)(2019).\\
3. J.Abelleira Fernandez et al., [LHeC Collaboration],
J.Phys.G{\bf39}, 075001(2012);
P.Agostini et al. [LHeC Collaboration and FCC-he Study Group ], CERN-ACC-Note-2020-0002, arXiv:2007.14491 [hep-ex] (2020).\\
4. F.Aaron et al., [H1 and ZEUS Collaborations], JHEP{\bf1001},
109(2010); J.Breitweg et al., [ZEUS Collaboration],
Phys.Lett.B{\bf407}, 432(1997);
 J.Breitweg et al., [ZEUS Collaboration], Phys.Lett.B{\bf487}, 53(2000).\\
5. A. Abada et al., [FCC Collaboration], Eur.Phys.J.C{\bf 79}, 474(2019); R.A.Khalek et al., SciPost Phys.{\bf7}, 051(2019).\\
6. A.Caldwell and M.Wing, Eur.Phys.J.C{\bf76}, 463(2016); A.Caldwell et al., arXiv[hep-ph]:1812.08110(2018).\\
7. L.P.Kaptari et al., JETP Lett.{\bf 109}, 281(2019); L.P.Kaptari et al., Phys.Rev.{\bf D99}, 096019(2019).\\
8. G.R.Boroun, Phys.Rev.C{\bf97}, 015206 (2018); B.Rezaei and
G.R.Boroun, Eur.Phys.J.A{\bf56}, 262(2020).\\
9. A.V.Kotikov and G.Parente, JHEP\textbf{85}, 17(1997);
A.V.Kotikov and G.Parente, Mod.Phys.Lett.A\textbf{12}, 963(1997);
A.M.Cooper-Sarkar et al., Z.Phys.C{\bf39}, 281(1988);
A.M.Cooper-Sarkar et al., Acta Phys.Pol.B{\bf34}, 2911(2003).\\
10. G.R.Boroun and B.Rezaei, Eur.Phys.J.C{\bf 72}, 2221(2012);
B.Rezaei and G.R.Boroun, Nucl.Phys.A{\bf 857}, 42(2011);
 G.R.Boroun, B.Rezaei and J.K.Sarma, Int.J.Mod.Phys.A{\bf 29}, 1450189(2014);
H.Khanpour, M.Goharipour and V.Guzey, Eur.Phys.J.C{\bf78}, 1(2018).\\
11. N.Baruah, N.M. Nath, J.K. Sarma , Int.J.Theor.Phys.{\bf 52}, 2464(2013).\\
12. M.M.Block et al., Phys.Rev.D{\bf89}, 094027 (2014).\\
13. E.A. Kuraev, L.N. Lipatov and V.S. Fadin, Phys. Lett.
B{\bf60}, 50(1975); Sov. Phys. JETP {\bf44}, 443(1976); Sov. Phys.
JETP {\bf45}, 199(1977); Ya. Ya. Balitsky and L.N. Lipatov, Sov.
J. Nucl. Phys.{\bf28}, 822(1978).\\
14. B.Badelek et al., J.Phys.G\textbf{22}, 815(1996); S. Catani and F. Hautmann, Nucl.Phys.B\textbf{427}, 475(1994).\\
15. B.I.Ermolaev and S.I.Troyan, Eur.Phys.J.C{80}, 98(2020);
M.Devee, R.Baishya and J.K.sarma, Eur.Phys.J.C{\bf72},
2036(2012).\\
16. G.R.Boroun and B.Rezaei, Eur.Phys.J.C{\bf73}, 2412(2013).\\
17. G.Altarelli and G.Martinelli, Phys.Lett.B\textbf{76}, 89(1978).\\
18.Yu.L.Dokshitzer, Sov.Phys.JETP {\textbf{46}}, 641(1977);
G.Altarelli and G.Parisi, Nucl.Phys.B \textbf{126}, 298(1977);
V.N.Gribov and L.N.Lipatov, Sov.J.Nucl.Phys. \textbf{15},
438(1972).\\
19. S.Moch, J.A.M.Vermaseren, A.Vogt, Phys.Lett.B \textbf{606},
123(2005).\\
20. W.L. van Neerven, A.Vogt, Phys.Lett.B \textbf{490}, 111(2000);
 A.Vogt, S.Moch, J.A.M.Vermaseren, Nucl.Phys.B \textbf{691}, 129(2004).\\
21. C.Lopez and F.J.Yndurain, Nucl.Phys.B{\bf171}, 231 (1980).\\
22. U.D$^{,}$Alesio et al., Eur.Phys.J.C{\bf9}, 601 (1999);
J.R.Cudell, A.Donnachie and
P.V.Landshoff, Phys.Lett.B{\bf448}, 281(1999); P.V.Landshoff, arXiv:hep-ph/0203084.\\
23. R.D.Ball et al., Eur.Phys.J.C{\bf76}, 383(2016).\\
24. M.Praszalowicz, Phys.Rev.Lett.{\bf106}, 142002(2011);
M.Praszalowicz and T.Stebel, JHEP.{\bf03},
090(2013).\\
25. B.Rezaei and G.R.Boroun, Eur.Phys.J.A{\bf55},
66(2019).\\
26. A.D.Martin, W.J.Stirling and R.G.Roberts, Phys.Rev.D{\bf50}, 6734(1994).\\
27. R.D.Ball and S.Forte, Phys.Lett.B{\bf335}, 77(1994); R.D.Ball and S.Forte, Phys.Lett.B{\bf336}, 77(1994).\\
28. D.Britzger et al., Phys. Rev. D {\bf100}, 114007 (2019).\\
29. C.Adloff et al. [H1 Collaboration], Phys.Lett.B{\bf
520}, 183(2001).\\
30. M. Froissart, Phys. Rev. {\bf123}, 1053 (1961).\\
31. M. M. Block and F. Halzen, Phys.Rev.Lett.{\bf107},
212002(2011); M. M. Block and F. Halzen, Phys. Rev. D{\bf70},
091901(2004).\\
32. G.R.Boroun, Lithuanian Journal of Physics {\bf48}, 121 (2008);
G.R.Boroun and B.Rezaei, Phys.Atom.Nucl.{\bf71}, 1077 (2008);
B.Rezaei and G.R.Boroun, Eur.Phys.J. A{\bf55}, 66(2019);
G.R.Boroun, Eur. Phys. J. Plus {\bf135}, 68 (2020);
 G.R.Boroun and B.Rezaei, Acta. Phys. Slovaca {\bf56}, 463 (2006).\\
33. J.Blumlein and H.Bottcher, arXiv[hep-ph]:0807.0248(2008);
M.R.Pelicer et al., Eur.Phys.J.C{79}, 9(2019).\\
34. A.M.Cooper-Sarkar, arXiv:1605.08577v1 [hep-ph] 27 May 2016; I.Abt et.al., arXiv:1604.02299v2 [hep-ph] 11 Oct 2016.\\
35. F.D. Aaron et al. [H1 Collaboration], Eur.Phys.J. C{\bf63}, 625(2009).\\
36. H.Khanpour, A.Mirjalili and S.Atashbar Tehrani Phys.Rev.C{\bf95}, 035201(2017).\\
37. G.R.Boroun and B.Rezaei, Nucl.Phys.A{\bf990}, 244(2019);
B.Rezaei
and G.R.Boroun, Nucl.Phys.A{\bf1006}, 122062(2021).\\
38. J.Lan et al., arXiv[nucl-th]:1907.01509 (2019);  J.Lan et al.,
arXiv[nucl-th]:1911.11676 (2019).\\
39. B.G. Shaikhatdenov, A.V. Kotikov, V.G. Krivokhizhin, G.
Parente, Phys. Rev. D {\bf81}, 034008(2010).\\
40. S. Chekanov et al. [ZEUS Collaboration], Eur. Phys. J. {\bf C21}, 443 (2001).\\
41. A.D.Martin et al., Phys.Letts.{\bf B604}, 61(2004).\\
42. A.Donnachie and P.V.Landshoff, Phys.Lett.B {\bf437}, 408(1998
); A.Donnachie and P.V.Landshoff, Phys.Lett.B {\bf550}, 160(2002 ).\\
43. K Golec-Biernat and A.M.Stasto, Phys.Rev.D {\bf80},
014006(2009).\\
44. Tie-Jiun Hou et al., Phys.Rev.D{\bf103}, 014013(2021).\\
45. S.Dulat et al., Phys. Rev. D{\bf93}, 033006 (2016).\\
46. M.Kuroda and D.Schildknecht, Phys.Lett. B{\bf618}, 84(2005); M.Kuroda and D.Schildknecht, Acta Phys.Polon. B{\bf37}, 835(2006);
M.Kuroda and D.Schildknecht, Phys.Lett. B{\bf670}, 129(2008); M.Kuroda and D.Schildknecht, Phys.Rev. D{\bf96}, 094013(2017).\\
47. D.Schildknecht and M.Tentyukov, arXiv[hep-ph]:0203028; M.Kuroda and D.Schildknecht, Phys.Rev. D{\bf85}, 094001(2012).\\
48. B.Rezaei and G.R.Boroun, Phys.Rev.C {\bf101}, 045202 (2020).\\
49. F.D. Aaron et al. [H1 Collaboration], phys.Lett.B\textbf{665},
139(2008).\\
50. H.Abromowicz et al. [ZEUS Collaboration],
Phys.Rev.D\textbf{9}, 072002(2014).\\
51. V.Andreev et al. [H1 Collaboration], Eur.Phys.J.C{\bf74}, 2814
(2014).\\

}
\begin{figure}
\includegraphics[width=1\textwidth]{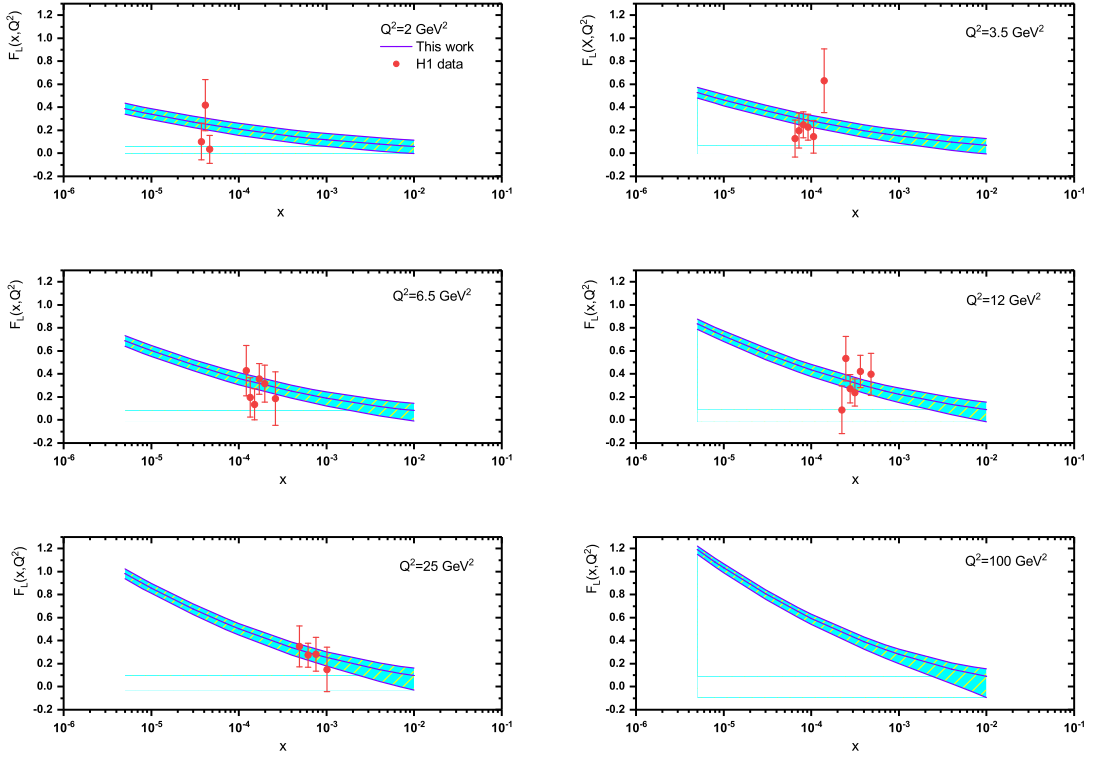}
\caption{The longitudinal structure function extracted in
comparison with the H1 data [1] as accompanied with total errors.
The results are presented  at fixed value of the inelasticity y
($y=0.49$). Combined H1 data for $E_{p}=460$ and
$575~\mathrm{GeV}$ gives $\sqrt{s}=225~\mathrm{GeV}$. The error
bands correspond to the uncertainty in the parameterization of
$F_{2}(x,Q^{2})$ in [12].}\label{Fig1}
\end{figure}
\begin{figure}
\includegraphics[width=1\textwidth]{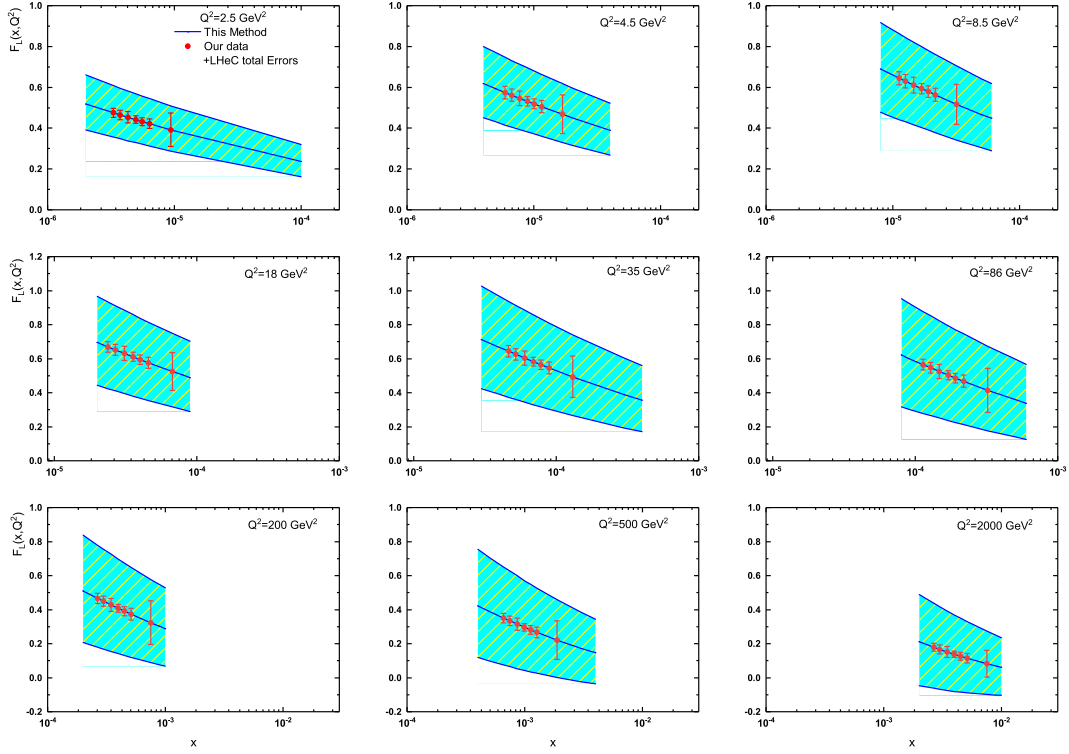}
\caption{The longitudinal structure function extracted at low $x$
in comparison with central our data as accompanied with the LHeC
total errors described in the CERN-ACC-Note-2020-0002 [3]. The
results are presented in the interval values of $Q^{2}$ (i.e.,
$2.5 ~ \mathrm{GeV}^{2}{\leq}Q^{2}{\leq}2000~\mathrm{GeV}^{2}$).
The error bands correspond to the uncertainty in the
parameterization of $F_{2}(x,Q^{2})$ in [12].}\label{Fig2}
\end{figure}
\begin{figure}
\includegraphics[width=1\textwidth]{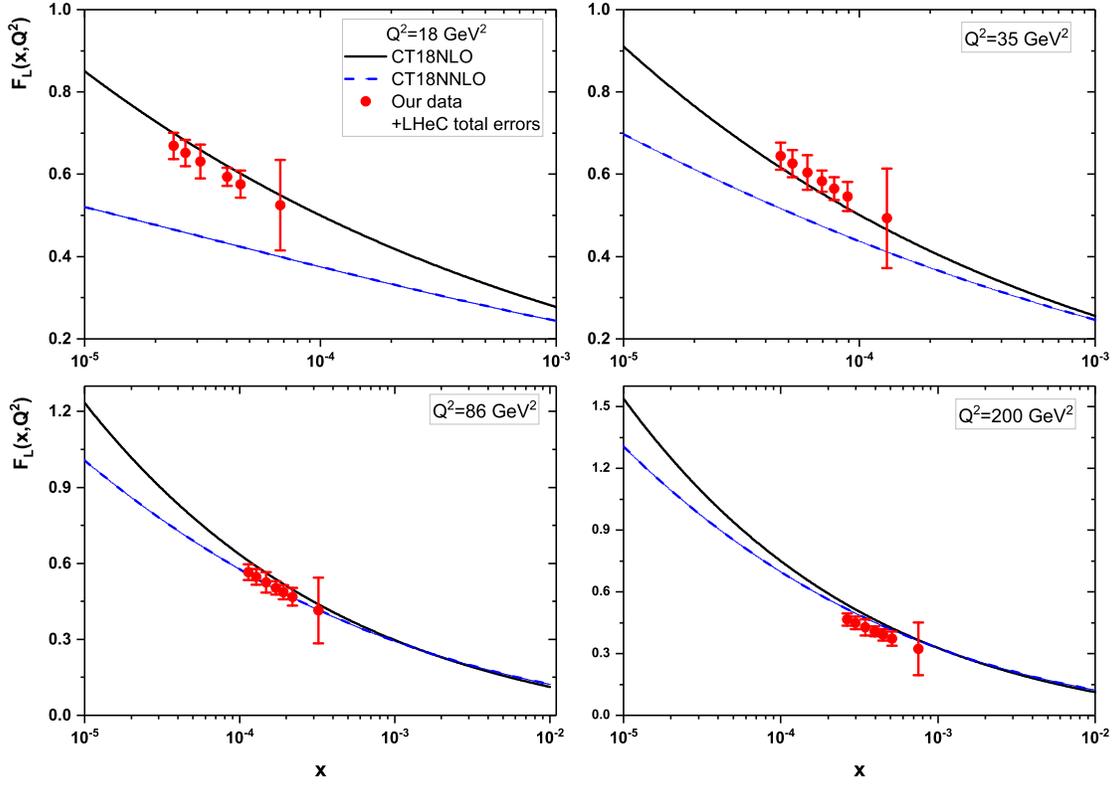}
\caption{The longitudinal structure function $F_{L}(x,Q^{2})$ with
respect to the LHeC simulated errors [3] in comparison with the
results of CT18 model [44] at $Q^{2}$ values $18$, $32$, $86$ and
$200~\mathrm{GeV}^{2}$.}\label{Fig3}
\end{figure}
\begin{figure}
\includegraphics[width=1\textwidth]{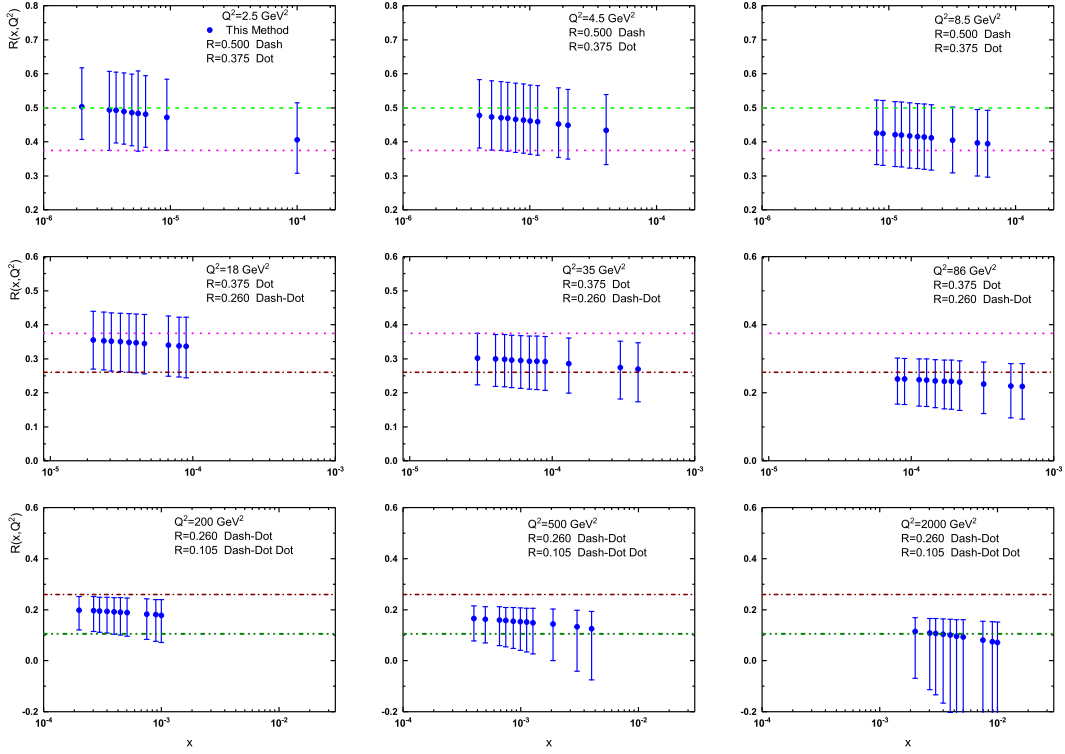}
\caption{Predictions of the ratio $R(x,Q^{2})$ over a wide range
of $Q^{2}$ values (i.e.,
$2.5{\leq}Q^{2}{\leq}2000~\mathrm{GeV}^{2}$) compared with other
results. Dash, Dot, Dash-Dot and Dash-Dot Dot lines are the
constant values in [45-51].}\label{Fig4}
\end{figure}
\begin{figure}
\includegraphics[width=.5\textwidth]{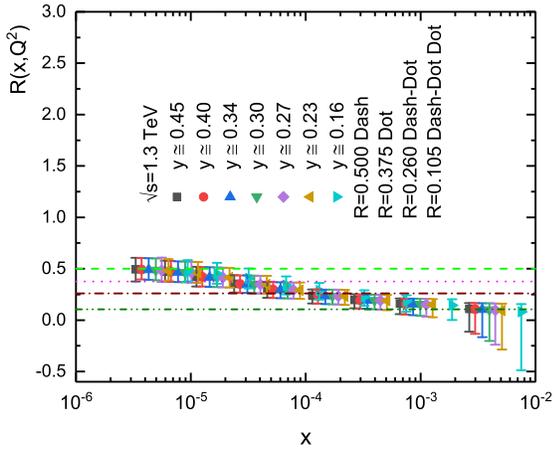}
\caption{Ratio $R(x,Q^{2})$ plotted as a function of $x$ variable
at $\sqrt{s}=1.3~\mathrm{TeV}$ in a wide range of $0.1{<}y{<}0.5$
compared with the dipole upper bounds [45-51].}\label{Fig5}
\end{figure}
\begin{figure}
\includegraphics[width=.5\textwidth]{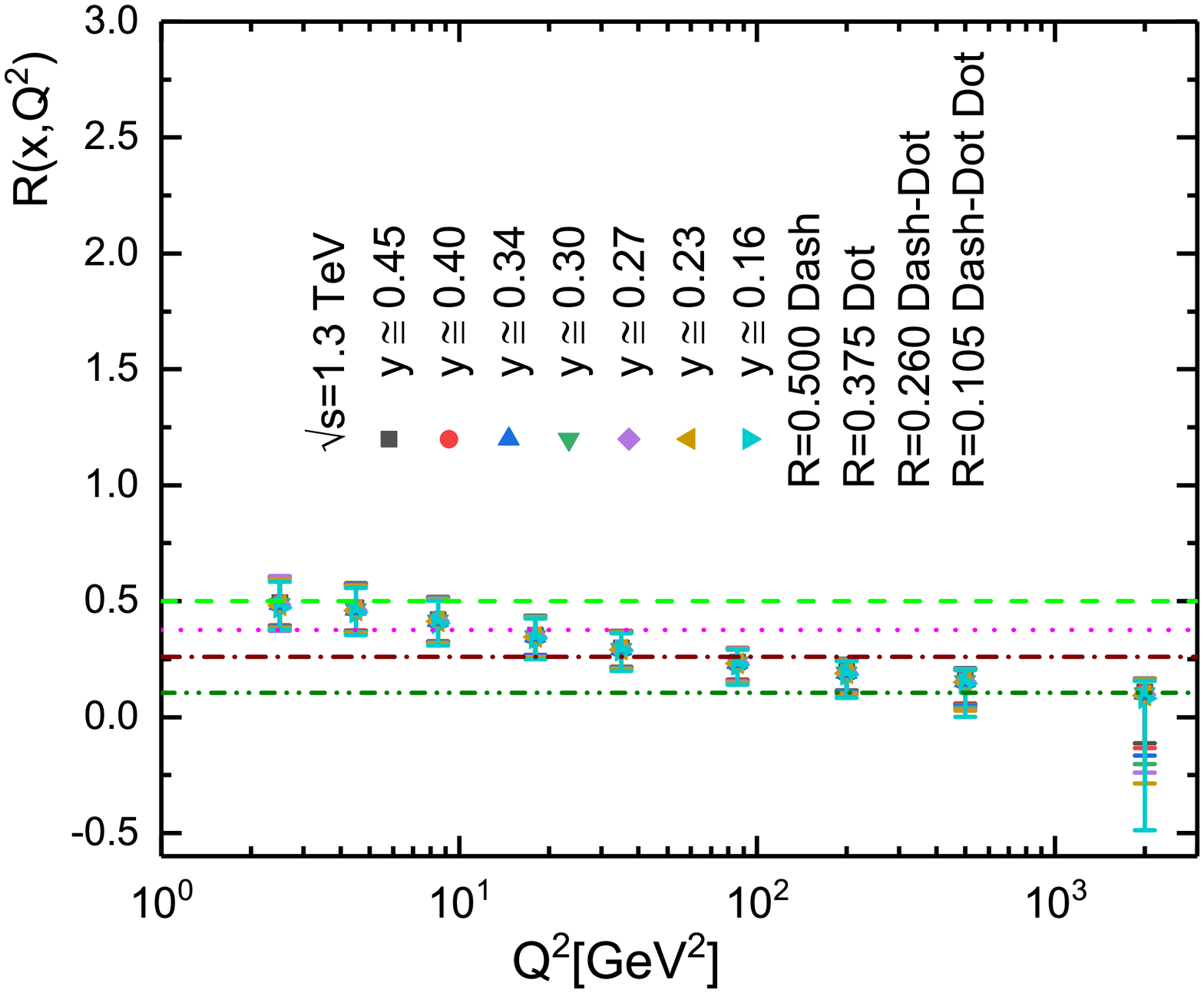}
\caption{ The same as Fig.5 for the ratio $R$ vs
$Q^{2}$.}\label{Fig6}
\end{figure}
\begin{figure}
\includegraphics[width=0.5\textwidth]{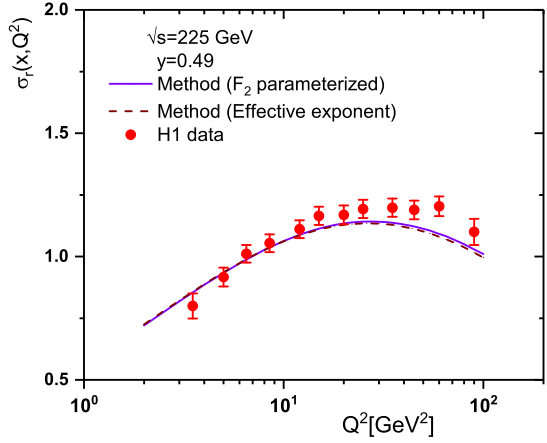}
\caption{The reduced cross section extracted in comparison with
the H1 data [1] as accompanied with total errors. The results are
presented  at fixed value of the inelasticity y ($y=0.49$).
Combined data for $E_{p}=460$ and $575~\mathrm{GeV}$ give
$\sqrt{s}=225~\mathrm{GeV}$. These results are based on two
methods: the parameterization of $F_{2}$ [12] and the effective
exponent [24].}\label{Fig7}
\end{figure}
\begin{figure}
\includegraphics[width=0.5\textwidth]{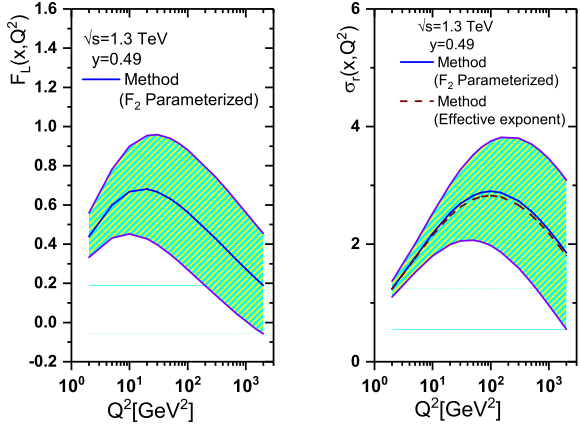}
\caption{The longitudinal structure function (Left: $F_{L}$) and
the reduced cross section (Right: $\sigma_{r}$) extracted at
$\sqrt{s}=1.3~\mathrm{TeV}$. The averaged value of inelasticity
has the value $y=0.49$. } \label{Fig8}
\end{figure}
\begin{figure}
\includegraphics[width=0.5\textwidth]{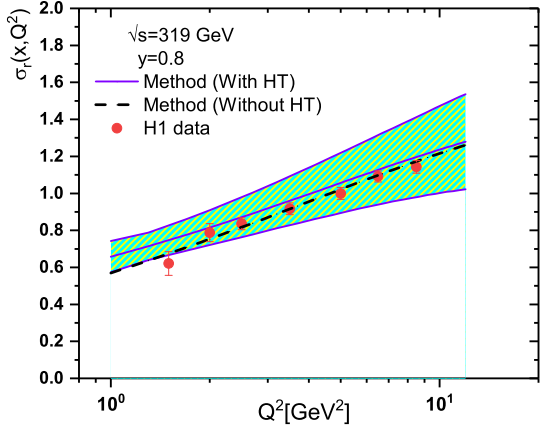}
\caption{The reduced cross section extracted in comparison with
the H1 data [1] at low $Q^{2}$ values with and without the HT
corrections. The results are presented at $y=0.8$ and
$\sqrt{s}=319~\mathrm{GeV}$. The error band corresponds to the
uncertainty in the parameterization of $F_{2}$ in [12].
}\label{Fig9}
\end{figure}
\begin{figure}
\includegraphics[width=0.5\textwidth]{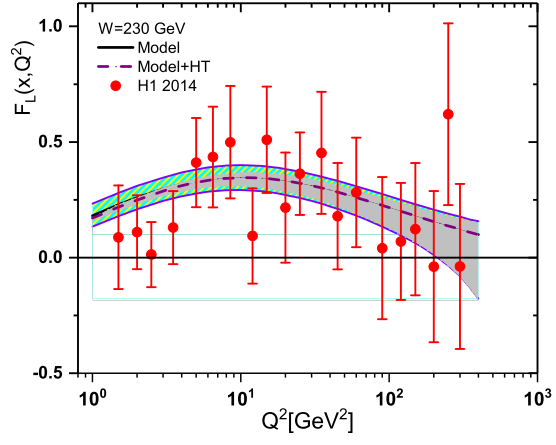}
\caption{The longitudinal structure function extracted with and
without HT corrections in comparison with the H1 data [51] as
accompanied with total errors. The results are presented at fixed
value of the invariant mass W ($W=230~\mathrm{GeV}$) at low values
of $x$. }\label{Fig10}
\end{figure}

\end{document}